\documentclass[11pt]{article}
\usepackage{amsfonts,amssymb, graphicx,a4,bm,bbm}
\usepackage{color}
\newtheorem{theo}{Theorem}[section]

\newtheorem{prop}[theo]{Proposition}

\newtheorem{rema}[theo]{Remark}
\newtheorem{prob}[theo]{Problem}
\definecolor{Red}{cmyk}{0,1,1,0}
\def\red{\color{Red}}

\let\m=\mu       \let\ph=\varphi

  \let\G=\Gamma \let\L=\Lambda

\let\\=\noindent
\def\ee{\end{equation}}
\def\be{\begin{equation}}
\def\vv{\vskip.2cm}
\def\0{\emptyset}

\def\PP{{\mathcal  P}}

\def\1{\rlap{\mbox{\small\rm 1}}\kern.15em 1}

\def\Rd{\mathbf{R}^{\mbox{\tiny\rm D}}}
\def\Rfp{\mathbf{R}^{\mbox{\tiny\rm FP}}}

\def\rd{{\rm R}^{\mbox{\tiny\rm D}}}
\def\rfp{{\rm R}^{\mbox{\tiny\rm FP}}}

\title{Abstract polymer gas. A simple inductive proof
of the Fern\'andez-Procacci criterion}
\author{
\\
Paula Mendes Soares Fialho\\
\\
\small{ Departamento de Matem\'atica UFMG}
\small{ 30161-970 - Belo Horizonte - MG
Brazil}
\\
\small{paulamendes26@hotmail.com}}

\begin{document}

\maketitle

\begin{abstract}
This note contains an alternative proof of  the Fern\'andez-Procacci criterion for the convergence of cluster expansion on the abstract polymer gas
via a simple inductive argument {\it a l\'a Dobrushin}.
\end{abstract}

\vskip.2cm
{\footnotesize

}
\vskip.5cm

\vskip.5cm
\def\EA{{E_{\cal A}}}
\def\red{\color{Red}}

\section{Introduction}\def\GG{{\cal G}}\def\EE{{\mathcal E}}

\\In statistical mechanics the abstract polymer gas is a discrete model defined by a triple $(\PP,\mathbf{w}, W)$ where $\PP$ is a countable set whose elements are called polymers,
$\mathbf{w}:\PP\to \mathbb{C}$ is a function   which associates to each  $x\in \PP$ a  complex number $w_x$
called the activity of the polymer $x$ and $W:\PP\times \PP\to \{0,1\}$ is a  function, called the {\it Boltzmann factor}, such that $W(x,x)=0$ for all $x\in \PP$ and $W(x,y)=W(y,x)$ for all $\{x,y\}\subset \PP$.
Usually the pair $\{x,y\}$ is  called {\it incompatible} when $W(x,y)=0$ and {\it compatible} when $W(x,y)=1$, then for each polymer $x \in \PP$ the pair $\{x,x\}$ is incompatible.

\\Let us denote by $\GG=(\PP, \EE)$ the simple graph with vertex set $\PP$ and edge set $\EE$ formed  by all the incompatible pairs $\{x,y\}\subset \PP$, i.e., such that $W(x,y)=0$. The neighborhood of a vertex $x$ of $\GG$ is the set $\G^*_\GG(x)=\{y\in \PP: W(x,y)=0\}$, in other words, $\G^*_\GG(x)$ is the set of all polymers in $\PP$ incompatible with $x$. Observe that $|\G^*_\GG(x)| \ge 1$, as each $x$ is incompatible with itself, note also that depending of the function $W$ the set  $\G^*_\GG(x)$ can be infinite.


\\Given $S \subseteq \PP$,  $S$ is said to be an {\it independent set} of $\GG$ if each pair $\{x,y\}\subset S$ does not belong to $\EE$, i.e., if $S$ is a set of pairwise compatible polymers. Let us denote by $I(\GG)$ the set formed by all {\it finite} independent sets of $\GG$.

\\Given a finite collection of polymers $\L\subset \PP$, the grand canonical partition function at ``finite volume"  $\L$
is given by  any one of the three expression below
\be\label{z0}
 Z_\L(\mathbf{w})= ~1+ \sum_{n\ge 1} {1\over n!}\sum_{(x_1,\dots, x_n)\in \L^n}w_{x_1}\dots w_{x_n}\prod_ {1\leq i<j\leq n}
W(x_i,x_j)
\ee
\be\label{z1}
=~\sum_{S\subseteq \L} \prod_{x\in S} w_x\prod_{\{x,y\}\subset S}W(x,y)~~~~~~~~~~~~~~~~~~~~~~~~~~~~~
\ee
\begin{equation}\label{z}
= ~\sum_{S \subseteq \L \atop S \in I(\GG)}{\prod_{x \in S}{w_x}},~~~~~~~~~~~~~~~~~~~~~~~~~~~~~~~~~~~~~~~~~~~~~~~
\end{equation}
with the convention here  and hereafter that the empty product is equal to one. When $S \subset \PP$ is not finite,  $Z_S(\mathbf{w})$ can still be seen as  a formal power series in the activities.

\\All “physical quantities” of the model can be deduced from the function $ Z_\L(\mathbf{w})$. In particular the ``pressure" of the system in the ``finite volume"   $\L$  is given by
\be\label{press}
P_\L(\mathbf{w})={1\over |\L|}\log Z_\L(\mathbf{w}),
\ee
and, given an  independent set $S\subset \L$, the  correlations are given by
\be\label{corr}
\phi_\L(\mathbf{w},S)\;=\;
\prod_{x\in S}w_x
 \frac{Z_{\L\backslash S}(\mathbf{w})}{Z_\L(\mathbf{w})}\;.
\ee
If $S\notin I(\GG)$ then $\phi_\L(\mathbf{w},S)=0$.
When the activity $w_x$ is real nonnegative for all $x\in \PP$ (shortly $\mathbf{w}\ge 0$), the correlation function $\phi_\L(\mathbf{w},S)$ coincides with the probability to have a configuration  of polymers in $\L$ containing  the set
$S$.

\\The model was originally proposed by Koteck\'y and Preiss \cite{KP} in 1985 as a generalization of a lattice polymer  model introduced by Gruber and Kunz \cite{GK} in 1968. Its relevance in statistical mechanics is very deep since it is  widely used, as a technical tool, to study a large number of physical systems, (e.g. spin and particle systems, percolative models and even field theories).

\\A key problem  for the  model  is
to find radii $\mathbf{R} = \{r_x\}_{x \in \PP}$ (with $r_x\ge 0$ for all $x\in \PP$) such that for $|w_x|<r_x$ for all $x\in \PP$
(shortly $|\mathbf{w}|<\mathbf{R}$), the pressure and correlation functions
are analytic functions with respect to activities {\it uniformly in $\L$}.
The current way to approach  this problem   is to
expand $\log Z_{\L}(\mathbf{w})$  as a Taylor series in the activities around $\mathbf{w}=\bm 0$
\be\label{may}
\log Z_{\L}(\mathbf{w})\;=\; \sum_{n=1}^{\infty}{1\over n!}
\sum_{(x_{1},\dots ,x_{n})\in\L^n}
\phi^{T}(x_1 ,\dots , x_n)\,w_{x_1}\dots w_{x_n},
\ee
and then to try to show that this series  is absolutely convergent in some polydisc  $|\mathbf{w}|\le \mathbf{R}$ for all $\L\in \PP$.
The  power series (\ref{may}) is known as {\it Mayer series} and its  coefficients $\phi^{T}(x_{1},\dots ,x_{n})$
admit a beautiful and long known explicit expression which depends  only on the  graph with  vertex set $\{x_1,x_2,\dots,x_n\}$ and edge set
$\bigl\{\{x_i,x_j\}\subset \{x_1,x_2,\dots,x_n\}: W(x_i,x_j)=0 \bigr\}$.   We do not need here
to give such explicit expression for $\phi^{T}(x_{1},\dots ,x_{n})$  (see, for instance, \cite{FP}, formula (2.4)), we
will just use the so-called  \emph{alternating-sign property}
\be\label{alte}
\bigl|\phi^{T}(x_{1},\dots ,x_{n})\bigr|\;=\; (-1)^{n-1}\phi^{T}(x_{1},\dots ,x_{n})\;.
\ee
Identity (\ref{alte}), which holds for all systems interacting via a nonnegative pair potential  (that is to say, $0\le W(x,y)\le 1$),  is known since the sixties (see e.g. \cite{Ru}). For a simple inductive
proof of (\ref{alte}) in the specific case of the abstract polymer gas see for instance ~\cite{Mi}.

\\Given  $x\in \L$, let us define
\begin{eqnarray}\label{theta}
\Theta^\L_{x}(\mathbf{w})& = & \log Z_{\L}(\mathbf{w})- \log Z_{\L\backslash\{x\}}(\mathbf{w})\label{dif}\\
&=& \displaystyle{\sum_{n=1}^{\infty}{1\over n!}
\mathfrak{}\sum_{(x_{1},\dots ,x_{n})\in\L^n\atop \exists i:~ x_i=x}
\phi^{T}(x_1 ,\dots , x_n)\;{w_{x_1}}\dots{w_{x_n}}}\;. \label{theta}
\end{eqnarray}
Then analyticity of $\Theta^\L_{x}(\mathbf{w})$ implies immediately analyticity of the pressure (\ref{press}) and correlations (\ref{corr}).
Indeed, setting $\L= \{x_1,\ldots,x_k\}$, we have
$$
\log Z_{\L}(\mathbf{w})\;=\; \Theta^\L_{x_1}(\mathbf{w})+\sum_{i=2}^k \Theta^{\L\backslash{\{x_1,\dots,x_{i-1}\}}}_{x_i}(\mathbf{w})
$$
and,
setting $S=\{x_1.\dots, x_p\}\subset \L$,
$$
\phi_{\L}\bigl(\mathbf{w},S\bigr)\;=\;\Big(\prod_{x\in S}w_x\Big)
\exp\Bigl(-\sum_{i=1}^{p-1}\Theta^{\L\backslash\{x_{i+1},\ldots,x_p\}}_{x_i}(\mathbf{w})-\Theta^\L_{x_p}(\mathbf{w})\Bigr)\;.
$$
It is useful to define the  following positive term series
\be\label{posi}
|\Theta|^{\L}_x(\mathbf{p})\;=\;
\sum_{n=1}^{\infty}{1\over n!}
\sum_{(x_{1},\dots ,x_{n})\in \L^n\atop \exists i:~ x_i=x}
|\phi^{T}(x_1 ,\dots , x_n)|\,p_{x_1}\cdots{p_{x_n}},
\ee
where $\mathbf{p}=\{p_x\}_{x \in \PP}$ are nonnegative numbers, and by the alternating sign property (\ref{alte}) we have that
\be\label{theta2}
|\Theta|^{\L}_x(\mathbf{p})=-\Theta^{\L}_x(-\mathbf{p}).
\ee 
Observe that if we can control the convergence of series (\ref{posi}) then the analyticity of the series (\ref{theta}) follows.
Indeed if  (\ref{posi}) converges then  the formal series (\ref{theta})  converges absolutely for any $\mathbf{w}$
in the polydisc $|\mathbf{w}|\le \mathbf{p}$. 


\\Therefore  the key problem  for the  model  can be written as
 the following.

\begin{prob}\label{p1}
  Find radii $\mathbf{R} = \{R_x\}_{x \in \PP}$, with $R_x\ge 0$ for all $x\in \PP$, such that for $|w_x|<R_x$ for all $x\in \PP$ (shortly $|\mathbf{w}|<\mathbf{R}$),
  $|\Theta|^{\L}_x(\mathbf{|\bm w|})$ is analytic and bounded with respect to activities {\it uniformly in $\L$}.
\end{prob}

\vv

\def\Rd{\mathbf{R}^{\mbox{\tiny\rm D}}}
\def\Rfp{\mathbf{R}^{\mbox{\tiny\rm FP}}}
\def\phd{\ph^{\mbox{\tiny\rm D}}}
\def\phfp{\ph^{\mbox{\tiny\rm FP}}}
\\Until 2007, the best  bound for the radii $\mathbf{R}$ was given by Dobrushin \cite{D}.
~In that year Fern\'andez and Procacci \cite{FP} improved the Dobrushin bound proving the following theorem.

\begin{theo}[Fern\'andez-Procacci criterion]\label{FPC}
Let $\bm\mu= \{\mu_x\}_{x \in \PP}$ be a collection of nonnegative numbers such that
 \be\label{fp}
 |w_x| \leq \rfp_x\equiv \frac{\mu_x}{\ph_x^*(\bm \mu)},~~~~~~~~~~\forall x\in \PP
 \ee
 with
$$
\ph_x^*(\bm \mu)= ~1+ \sum_{n\ge 1} {1\over n!}\sum_{(x_1,\dots, x_n)\in \PP^n}\m_{x_1}\dots \m_{x_n}\prod_{i=1}^n[1-W(x,x_i)]
\prod_ {1\leq i<j\leq n}W(x_i,x_j)
$$
\be\label{zzz}
~=~\sum_{S\subset \PP\atop S~{\rm finite}}\prod_{y\in S}\{\m_{y}[1-W(x,y)]\}
\prod_{\{y,z\}\subset S}W(y,z),~~~~~~~~~~~~~~~~~~~~~~~~~
\ee
then  the  series  $|{\Theta}|^{\L}_x(|\bm w|)$, defined in (\ref{posi}) is convergent and furthermore
\be\label{lovaz}
|{\Theta}|^{\L}_x
(\bm |\bm w|)\;\le\: \ln(1+\m_x).
\ee
\end{theo}

\vskip.2cm

\begin{rema}
   It is easy to see that each parcel of (\ref{zzz}) will not vanish if, and only if, $S$ is an independent set in $\GG$, such that $S \subseteq \G^*_\GG(x)$. Then we will make an abuse of notation and write $\ph_x^*(\bm \mu) \equiv~Z_{\G^*_\GG(x)}(\bm \mu)$ (remember that the partition function $Z_{\L}$ was defined for finite sets $\L$, while $\G^*_\GG(x)$ can be infinite depending on $W$).
\end{rema}

\begin{rema}\label{rc}
  Note that $\ph_x^*(\bm \mu)$ is in general a positive term powers series in $\bm \mu$. So,  to make sure  that the criterion  (\ref{fp}) is non trivial, the numbers $\bm\mu$ must be chosen such that $\ph_x^*(\bm \mu)<+\infty$. Observe that $\ph_x^*(\bm \mu) < {\exp}(\sum_{y \in \G^*_\GG(x)}\mu_{y})$, therefore  $\ph_x^*(\bm \mu)$ is finite for any choice of $\bm \mu$ such that $ \sum_{y \in \G^*_\GG(x)}\mu_{y}$ is finite. 
\end{rema}
\def\VU{{\mathbb{V}}}
\begin{rema}\label{subset} A typical realization of the abstract polymer gas that appear in most of the uses of the cluster expansion in
statistical mechanics is the so-called
subset gas. Its definition requires a countable set $\VU$ that acts as an underlying ``space''.  Polymers are then simply defined as the finite non empty subsets of  $\VU$, i.e.
$$
\PP_\VU=\{x\subset \VU : 0< |x|<\infty\}
$$
and of course a family of activities $\bm w=\{w_x\in \mathbb{C}\}_{x\in \PP_\VU}$ is associated to $\PP_\VU$. In this situation the physical pressure at the ``real finite volume" $\L\subset \VU$ is defined as
$ P_\L={1\over |\L|}\ln Z_{\PP_\L}(\bm w)$, where $\PP_\L=\{ x\subset \VU : x\subset \L\}$. It is easy to show that (\ref{lovaz}) implies that the pressure $P_\L$ is uniformly bounded in $\L$ by choosing the numbers $\bm \m$
in such a way that $$\sup_{v\in \VU}\sum_{x\in \PP_\VU; \atop v\in x}\m_x=K<\infty.$$ We discuss more details about the pressure $ P_\L$ in the Appendix.
\end{rema}

\\The Dobrushin criterion admits  a nearly  identical  statement with the sole difference that condition (\ref{fp})
is replaced by
 \be\label{do}
 |w_x| \leq \rd_x\equiv  \frac{\mu_x}{\phd_x(\bm \mu)},~~~~~~~~~~\forall x\in \PP
 \ee
 with
$$
{\phd_x(\bm\mu)}=\sum_{S\subset \PP\atop S~{\rm finite}}\prod_{y\in S}\{\mu_{y}[1-W(x,y)]\}.
$$
Note that again $S$ must be a subset of $ \G^*_\GG(x)$, however now it is not necessary to be an independent set. Therefore $\phd_x(\bm\mu)\ge \ph^*_x(\bm \mu) $, for any  $x, \bm \mu$, and hence the bound $\Rfp$ on radii $\mathbf{R}$ given by the Fern\'andez-Procacci criterion (\ref{fp})
 is always greater than the
bound $\Rd$ on the same radii  given by  the Dobrushin criterion (\ref{do}). On the other hand, while the Dobrushin criterion can be proved through a simple and straightforward inductive argument, the proof of the Fern\'andez-Procacci criterion given in \cite{FP} involves a heavy combinatorial machinery and in particular makes use of  cluster expansion and tree-graph inequalities.

\\The objective of this note is to provide an alternative proof of Theorem \ref{FPC} based on a simple inductive argument. This new proof has been inspired by the connection between the abstract polymer model
and the Lov\'asz Local Lemma in combinatorics {(see \cite{AS} and references therein)}  pointed out by Scott and Sokal \cite{SS} and further developed by Bissacot et al. \cite{BFPS}. In this regards  we are greatly indebted to the work done by Harvey and Vondrak in \cite{HV},  where a generalization of the Moser-Tardos algorithmic version \cite{MT}  of the Lov\'asz Local Lemma is given, and notably to their Section 5.7 where the proof of Theorem \ref{FPC} presented below is somehow implicitly outlined.

\section{Proof of Theorem \ref{FPC}}

We will start by recalling two important properties of the partition function $
Z_{\L}(\mathbf{w})$  which follows straightforwardly from the definition (\ref{z})  (see e.g. \cite{SS}).
\begin{enumerate}
    \item
{\it Fundamental identity:} let $\L\subset \PP$ finite and $x\in \L$, then
    \be\label{fuid}
    Z_\L(\mathbf{w})= Z_{\L \setminus \{x\}}(\mathbf{w}) + w_x Z_{\L \setminus \Gamma_\GG^{*}(x)}(\mathbf{ w}).
    \ee
    \item {\it Log-subadditivity:} let $S, T \subseteq \PP$  and let $\bm\m=\{\m_x\}_{x\in \PP}$ with $\m_x\ge 0$ for all $x \in \PP$,  then
    $$Z_{S \cup T}(\bm\mu) \leq Z_S(\bm\mu) Z_T(\bm{\mu}).$$
\end{enumerate}


\\Given  nonnegative  numbers $\mathbf{p}=\{p_x\}_{ x \in \PP}$, for every finite set $S \subseteq \PP$ we set shortly

  \be\label{qs}
  Q_S(\mathbf{p})= Z_S(-\mathbf{p})
  \ee
and  observe  that via  the fundamental identity we have, for all $x \in S$
\be\label{qfu}
Q_S(\mathbf{p})=  Q_{S \setminus \{x\}}(\mathbf{p})
-p_x  Q_{S \setminus \Gamma^{*}_\GG(x)}(\mathbf{p}).
\ee
Recall that
${Z_{\Gamma^{*}_\GG(x)}(\bm\mu)}= \ph_x^{*}(\bm\mu)$.
The following proposition is a development of Harvey and Vondr\'ak (\cite{HV}, Lemma 5.41).
\begin{prop}\label{prop}
Let $\bm\mu\equiv \{\mu_x\}_{x \in \PP}$ and $\mathbf{p}\equiv \{p_x\}_{x \in \PP}$ be nonnegative numbers  such that
   \be\label{ffpp}
   p_x \le \rfp_x=\frac{\mu_x}{Z_{\Gamma^{*}_\GG(x)}(\bm\mu)},~~~~\forall x\in \PP.
   \ee
Given a finite set $\L\subset \PP$, let $S \subseteq \L$ and let $S^c=\L\setminus S$. Then
  \begin{equation}\label{Q}
    \frac{Q_S(\mathbf{p})}{Q_{S\setminus \{x\}}(\mathbf{p})} \ge \frac{Z_{S^{c}}(\bm \mu)}{Z_{(S \setminus \{x\})^c}(\bm \mu)}, ~~~
    \mbox{for all $x \in S$}.
  \end{equation}
  .
\end{prop}

\\{\bf Proof}.
Let us use induction on $|S|$. For $S=\{x\}$, observe that
$$
\frac{Q_{\{x\}}(\mathbf{p})}{Q_\emptyset(\mathbf{p})}= \frac{1-p_x}{1}=1-p_x.
$$
By other hand, $S^{c}= \L \setminus \{x\}$ and $(S \setminus \{x\})^{c} = \L$, then

\begin{eqnarray}
  Z_\L(\bm\mu) &=& Z_{\L \setminus \{x\}}(\bm\mu) + \mu_x Z_{\L \setminus \Gamma_\GG^{*}(x)}(\bm\mu) \label{1} \\
          &\geq& Z_{\L \setminus \{x\}}(\bm\mu) + p_x Z_{\G^*_\GG(x)}(\bm \mu)  Z_{\L \setminus \Gamma^{*}_\GG(x)}(\bm\mu)\label{2}  \\
          &\geq& Z_{\L \setminus \{x\}}(\bm\mu) + p_x Z_{\L\cup \G^*_\GG(x)}(\bm\mu)\label{3}\\
          &\geq& Z_{\L \setminus \{x\}}(\bm\mu) + p_x Z_{\L}(\bm\mu)\label{4},
\end{eqnarray}
where Equality (\ref{1}) is due to fundamental identity, Inequality (\ref{2}) is due to (\ref{ffpp}), Inequality (\ref{3}) is due log-subadditivity
and Inequality (\ref{4}) follows trivially form the definition of partition function. By Inequality (\ref{4}) we have
$$\frac{Z_{\L \setminus \{x\}}(\bm\mu)}{Z_\L(\bm \mu)} \le 1-p_x,$$
therefore
$$\frac{Q_{\{x\}}(\mathbf{p})}{Q_\emptyset(\mathbf{p})}\ge \frac{Z_{\L \setminus \{x\}}(\bm\mu)}{Z_\L(\bm \mu))}.$$

\\Now suppose that for any set $T\subset \L$ with cardinality less than $n$ and for any $x\in T$ we have that
$$ \frac{Q_T(\mathbf{p})}{Q_{T\setminus \{x\}}(\mathbf{p})}\ge \frac{Z_{T^{c}}(\bm \mu)}{Z_{(T \setminus \{x\})^c}(\bm \mu)}.$$
So we will prove that (\ref{Q}) holds for a set $S\subset \L$ such that $|S|=n$ and for any polymer $x\subset S$. By (\ref{qfu}) we have

\begin{equation}\label{hi}
 \frac{Q_S(\mathbf{p})}{Q_{S\setminus \{x\}}(\mathbf{p})} =\frac{Q_{S \setminus \{x\}}(\mathbf{p}) -p_x Q_{S \setminus \Gamma^{*}_\GG(x)}(\mathbf{p})}{Q_{S\setminus \{x\}}(\mathbf{p})}= 1 - p_{x} \frac{Q_{S \setminus \Gamma^{*}_\GG(x)}(\mathbf{p})}{Q_{S\setminus \{x\}}(\mathbf{p})}.
\end{equation}

\\Observe now that $S \setminus \Gamma^{*}_\GG(x)= S \setminus (\Gamma^{*}_\GG(x)\cap S)$. Suppose that $|\Gamma^{*}_{\GG}(x)\cap S|=m+1$ and let us write  $\Gamma^{*}_{\GG}(x)\cap S=\{x, x_1, \dots, x_{m}\}$, then
$$\frac{Q_{S\setminus \{x\}}(\mathbf{p})}{Q_{S \setminus \Gamma^{*}_\GG(x)}(\mathbf{p})}=
\frac{Q_{S \setminus \{x\}}(\mathbf{p})}{Q_{S \setminus \{x \cup x_1\}}(\mathbf{p})}
\frac{Q_{S \setminus \{x \cup x_1\}}(\mathbf{p})}{Q_{S \setminus \{x \cup x_1 \cup x_2\}}(\mathbf{p})}
\cdots \frac{Q_{S \setminus \{x \cup x_1 \cup x_2 \cup \dots \cup x_{m-1}\}}(\mathbf{p})}{Q_{S \setminus \Gamma^{*}_\GG(x)}(\mathbf{p})}$$
and as each parcel satisfies the induction hypotheses, we have that
$$\frac{Q_{S\setminus \{x\}}(\mathbf{p})}{Q_{S \setminus \Gamma^{*}_\GG(x)}(\mathbf{p})}
\ge
\frac{Z_{(S \setminus\{x\})^c}(\bm\mu)}{Z_{(S \setminus\{x \cup x_1\})^c}(\bm\mu)} \frac{Z_{(S \setminus\{x \cup x_1 \})^c}(\bm\mu)}
{Z_{(S \setminus\{x \cup x_1 \cup x_2 \})^c}(\bm\mu)} \dots \frac{Z_{(S \setminus\{x \cup x_1 \cup x_2 \cup \dots \cup x_{m-1}\})^c}(\bm\mu)}{Z_{(S \setminus \Gamma^{*}_\GG(x))^c}(\bm\mu)}
$$
$$
=\frac{Z_{(S \setminus\{x\})^c}(\bm\mu)}{Z_{(S \setminus \Gamma^{*}_\GG(x))^c}(\bm\mu)}.~~~~~~~~~~~~~~~~~~~~~~~~~~~~~~~~~~~~~~~~~~~~~~$$
Then by Equation (\ref{hi})
\begin{eqnarray}
\nonumber \frac{Q_S(\mathbf{p})}{Q_{S\setminus \{x\}}(\mathbf{p})}  &=&1 - p_{x} \frac{Q_{S \setminus \Gamma^{*}_\GG(x)}(\mathbf{p})}{Q_{S\setminus \{x\}}(\mathbf{p})} \\
\nonumber                                  &\ge& 1 -p_x \frac{Z_{(S \setminus \Gamma_\GG^{*}(x))^c}(\bm \mu)}{Z_{(S \setminus\{x\})^c}(\bm\mu)} \\
\nonumber                                  &=& 1 -p_x \frac{Z_{S^c \cup (\Gamma^{*}_\GG(x)\cap S)}(\bm\mu)}{Z_{S^c \cup \{x\}}(\bm\mu)} \\
                                           &\ge& 1 - \frac{Z_{S^c \cup \{x\}}(\bm\mu)- Z_{S^c}(\bm\mu)}{Z_{S^c \cup \{x\}}(\bm\mu)} \label{eq}\\
\nonumber                                  &=& \frac{Z_{S^c}(\bm\mu)}{Z_{S^c \cup \{x\}}(\bm\mu)} \\
\nonumber                                   &=& \frac{Z_{S^c}(\bm\mu)}{Z_{(S \setminus \{x\})^c}(\bm\mu)}
\end{eqnarray}
where Inequality (\ref{eq}) is due to

\begin{eqnarray}
\nonumber  Z_{S^c \cup \{x\}}(\bm\mu) &=& Z_{S^c}(\bm\mu) + \mu_x Z_{S^c \setminus \Gamma^{*}_\GG(x)}(\bm\mu) \\
\nonumber          &\geq& Z_{S^c}(\bm\mu) + p_x Z_{\Gamma^{*}_\GG(x)}(\bm\mu) Z_{S^c \setminus \Gamma^{*}_\GG(x)}(\bm\mu)  \\
\nonumber          &\geq&  Z_{S^c}(\bm\mu)  + p_x Z_{S^c \cup \Gamma^{*}_\GG(x) }(\bm\mu)  \\
\nonumber          &\geq&  Z_{S^c}(\bm\mu)  + p_x Z_{S^c \cup (\Gamma^{*}_\GG(x)\cap S)}(\bm\mu).
\end{eqnarray}

~~~~~~~~~~~~~~~~~~~~~~~~~~~~~~~~~~~~~~~~~~~~~~~~~~~~~~~~~~~~~~~~~~~~~~~~~~~~~~~~~~~~~~~~~~~~~~~~~~~~~~~~~~~~~~~~$\Box$
\vv
\\{To conclude the } proof of Theorem \ref{FPC},
we can now apply  Proposition \ref{prop} taking   $S=\L$. Then, for any finite $\L\subset \PP$, for all $x\in \L$ and for all $\mathbf{p}\le \Rfp$ { we have}
$$
\frac{Z_\L(-\mathbf{p})}{Z_{\L\setminus \{x\}}(-\mathbf{p})}=
\frac{Q_\L(\mathbf{p})}{Q_{\L\setminus \{x\}}(\mathbf{p})}
\ge \frac{Z_{\emptyset}(\bm \mu)}{Z_{ \{x\}}(\bm \mu)} = \frac{1}{1+\m_x},
$$
and therefore, for all $\L$ finite and all $x\in \L$, we have that
$$
\frac{Z_{\L\setminus{x}}(-\mathbf{p})}{Z_{\L}(-\mathbf{p})}\le 1+\m_x,
$$
and hence, by (\ref{theta2}),
$$
|\Theta|^\L_x(\bm p)= -\Theta_x^\L(-\bm p)\le \ln(1+\m_x),
$$
which concludes the proof of Theorem \ref{FPC}.

\section*{Appendix}
Given a subset gas with underlying space $\VU$ and set of polymers $\PP_\VU=\{x\subset \VU:\;|x|<+\infty\}$  with activities $\bm w=\{w_x\}_{x\in \PP_\VU}$, let
 $\L\subset \VU$ be a finite set and let   $\PP_\L=\{x\subset \VU:\;x\subset \L\}$. The pressure of the system is defined as the following function
$$
P_\L(\bm w)={1\over |\L|}\ln Z_{\PP_\L}(\bm w)={1\over |\L|}\sum_{n=1}^\infty {1\over n!} \sum_{(x_{1},\dots ,x_{n})\in\PP_\L^n}
\phi^{T}(x_1 ,\dots , x_n)\,w_{x_1}\dots w_{x_n}.
$$
Suppose that
(\ref{fp}) holds with $\bm \mu$ chosen in such a way that
$$
\sup_{v\in \VU}\sum_{x \in \PP_\VU:\atop v\in x}\m_x<K
$$
then let us show that $P_\L(\bm w)$ is absolutely convergent and  $|P_\L(\bm w)|\le K$. Indeed, recalling that (\ref{fp}) implies (\ref{lovaz}) and recalling also definition
(\ref{posi}), we have
\begin{eqnarray*}
  |\ln Z_{\PP_\L}(\bm w)| &\le& \sum_{n=1}^\infty {1\over n!} \sum_{(x_{1},\dots ,x_{n})\in\PP_\L^n} |\phi^{T}(x_1 ,\dots , x_n)|\,|w_{x_1}|\dots |w_{x_n}| \\
                          &\le&  \sum_{v\in \L}\sum_{x\in \PP_\L; \atop v \in x}\sum_{n=1}^\infty {1\over n!} \sum_{(x_{1},\dots ,x_{n})\in\PP_\L^n\atop \exists i: x_i=x} |\phi^{T}(x_1 ,\dots , x_n)|\,|w_{x_1}|\dots |w_{x_n}| \\
                           &=& \sum_{v\in \L}\sum_{x\in \PP_\L; \atop v \in x}|\Theta|^\L_x(|\bm w|) \\
                           &\le& \sum_{v\in \L}\sum_{x\in \PP_\L; \atop v \in x}\ln(1+\m_x) \\
                           &\le& \sum_{v\in \L}\sum_{x\in \PP_\VU; \atop v \in x}\m_x\\
                           &\le& K |\L|.
\end{eqnarray*}

\vv
\\{\it Acknowledgements.} I would like to thank Aldo Procacci and Roberto Fernández for some helpful
comments on this note. The author is supported by Conselho Nacional de Desenvolvimento Cient\'ifico e Tecnol\'ogico (CNPq).

\appendix

\end{document}